\documentclass[twocolumn,nofootinbib,preprintnumbers,amsmath,amssymb]{revtex4}
\usepackage{graphicx,color}
\usepackage{bm}%
\setcitestyle{super}
\usepackage{bm}

\newcommand{\rr}{{\bf r}}
\newcommand{\rrh}{{\bm\rho}}

\newcommand{\EE}{{\cal E}}
\newcommand{\Ee}{{\bf E}}
\newcommand{\BB}{{\bf B}}

\newcommand{\Gg}{{\sf G}}

\newcommand{\PP}{{\bf P}}

\newcommand{\be}{\begin{equation}}
\newcommand{\ee}{\end{equation}}
\newcommand{\ba}{\begin{eqnarray}}
\newcommand{\ea}{\end{eqnarray}}
\newcommand{\bse}{\begin{subequations}}
\newcommand{\ese}{\end{subequations}}
\newcommand{\beq}{\begin{eqnarray}}
\newcommand{\eeq}{\end{eqnarray}}

\begin{document}
\title{What is the right theory for Anderson localization of light?}
\author{Walter Schirmacher$^{1,2,3}$, Behnam Abaie$^4$, Arash Mafi$^4$, Giancarlo Ruocco$^{1,2}$
\& Marco Leonetti$^{1,5}$}
\affiliation{%
$^1$Center for Life Nano science @ Sapienza, Isituto Italiano di Tecnologia, Viale Regina
Elena, 291, I-00161 Roma, Italia}
\affiliation{%
$^2$Dipartimento di Fisica, Universit\'{a} ``La Sapienza'', Piazzale Aldo Moro, 5, I-00185 Roma, Italia,}
\affiliation{%
$^3$Institut f\"ur Physik, Universit\"at Mainz, Staudinger Weg 7,
D-55099 Mainz, Germany,}
\affiliation{%
$^4$Department of Physics and Astronomy and Center for High Technology materials, University
of New Mexico, Albuquerque, NM 87131, USA,}
\affiliation{%
$^5$CNR NANOTEC, Istituto di Nanotechologia, I-73100 Lecce, Italia%
}

\begin{abstract}
Anderson localization of light is traditionally described in analogy
to electrons in a random potential. Within this description the
disorder strength -- and hence the localization characteristics --
depends strongly on the wavelength of the incident light.
In an alternative description in analogy to sound waves in
a material with spatially fluctuating elastic moduli this is not the case.
Here, we report on an {\it experimentum crucis} in order
to investigate the validity of the two conflicting theories
using transverse-localized optical devices. We do not find
any dependence of the observed localization radii on the light wavelength.
We conclude that the modulus-type description is the correct one
and not the potential-type one.
We corroborate this by showing that in the derivation of the
traditional, potential-type theory a term in the wave equation has been tacititly
neglected. In our new modulus-type theory the wave equation is exact.
We check the consistency of the new theory with our data
using a field-theoretical approach (nonlinear sigma model).
\end{abstract}
\maketitle
Anderson localization,
i.e. the possibility of an arrest of the motion
of an electronic wave packet
in a disordered environment, has fascinated
researchers since the appearance of Anderson's 1958 article
\cite{anderson58}.
In 1979 it became clear \cite{abrahams79}
that this phenomenon is due to destructive interference of
recurrent scattering paths and led - via the one-parameter scaling hypothesis -
to the conclusion that in disordered one- and two-dimensional systems the
waves are always localized. This scaling theory was put on solid theoretical grounds by relating the
Anderson scenario to the nonlinear sigma model of planar
magnetism \cite{wegner79,schafer80,mckane81}
and by the self-consistent diagrammatical
localization theory
\cite{vollhardt80,vollhardt82,wolfle10}.
It became clear that this transition should
exist
in any physical system governed
by a wave equation with disorder \cite{evers08}.

Anderson localization
has gained much attention recently in wave optics%
\cite{john91}
due to a large
number of possible applications reaching from solar cells to
endoscopic fibres%
\cite{soukoulis01,wiersma13,segev13}.

In the description of possible localization of light
by means of the nonlinear-sigma-model theory John
\cite{john84,john85,john87} adopted the same
structure of a classical wave equation with disorder
as in his sound-wave theory
\cite{john83,john83a}, namely a fluctuating coefficient
of the double time derivative.
In the time-Fourier-transformed
version of the wave equation
this version had the attractive feature that a one-to-one
mapping to the Schr\"odinger equation of an electron in
a random potential could be established.
So most of the results
of the theory for the electronic Anderson localization could be taken over
\cite{john84,john85,john87,kroha83,sheng06,wolfle10}. We call this approach the
``potential-type'' description (PT).

On the other hand, in an alternative formulation,
used successfully for the vibrational anomalies
in glasses \cite{schirm04,schirm06,schirm07}
the disorder enters the
coefficient of the spatial derivatives,
which in the case of sound waves is the
elastic modulus, in the case of electromagnetic
waves the
the dielectric modulus $1/\epsilon(\rr)$. We call this
the
``modulus-type'' description (MT).

While the existence of Anderson localization of light in 3-dimensional
optical materials is still under debate \cite{skipetrov16,sperling16},
in optical systems with restricted dimensionality
one has nowadays evidence for Anderson localization,
in particular
in optical fibers with transverse disorder
\cite{schwartz07,karbasi12,segev13}.
The possibility of observing transverse localization of light
in optical fibers, which are translation invariant along the fiber
axis but exhibit disorder in one or two of the transverse directions,
was already predicted some time ago
\cite{abdullaev80,deraedt89}.

In fibers composed of microfibers
with different dielectric constants the presence of transverse
localization leads to the existence of channels with
the diameter of the transverse localization length,
which transmit light like in a micro-waveguide.
As the localized modes have been proven to be of
single-mode character \cite{ruocco16},
such fibers are extremely useful for
transfer of multiple information, e.g for endoscopy.

The theoretical description
of transverse Anderson localization in fibers with
transverse disorder \cite{deraedt89,karbasi12,karbasi12a} followed
the potential-type
approach.
This description results in a rather strong dependence of the localization
lengths
on the wavelength of the applied light
\cite{karbasi12a}.

In the present contribution we show that this description is
not consistent
with our experimental observation.
We have measured the localization lengths of fibers
with transverse disorder as a function of the light wavelength
and do not find any change with wavelength.
Motivated by this observation, we adopted the modulus-type
approach do disorder and found perfect consistency with the
experiments. We conclude
that the modulus-type description is the correct
theory for Anderson localization of light.
A further argument against the
potential-type approach is that within this model the fibers are
opaque in the longitudinal ($z$) direction. Due to the Rayleigh-scattering
mechanism this is not the case in the modulus-type theory,
in which the samples are transparent in the $z$ direction.

In the frequency regime (with frequency variable
\mbox{$\omega =2\pi c_0/\lambda\sqrt{\langle\epsilon\rangle}=
c_0k_0/\sqrt{\langle\epsilon\rangle}$,} where
$c_0$ is the vaccum light speed,
$\lambda$ is the wavelength
$k_0=2\pi/\lambda$ is the wavenumber
and $\langle\epsilon\rangle$
the average permittivity)
the two conflicting stochastic
wave equations, which are both considered to
be derived from Maxwell's equations
with inhomogeneous permittivity,
take the form

\ba\label{potential1}
\big[\tilde\epsilon(\rr)k_0^2+\nabla^2\big]E_\alpha(\rr,\omega)&=&0\qquad\qquad\mbox{PT}\nonumber\\
\big[k_0^2+\nabla\frac{1}{\tilde\epsilon(\rr)}\nabla\big]B_\alpha(\rr,\omega)&=&0\qquad\qquad\mbox{MT}
\ea
$E_\alpha, B_\alpha$ are components
of the electric, magnetic field, resp. and
$\tilde\epsilon(\rr)=\epsilon(\rr)/\langle\epsilon\rangle$ denotes the relative
fluctuations of the dielecric constant.
In what follows we disregard the vector character of the electromagnetic
fields.

In the case of transverse disorder $\epsilon$ fluctuates only in
the $x,y$
direction, so one can perform a Fourier
transform with respect to the $z$ direction
($\partial/\partial z\rightarrow ik_z$)
to obtain
\ba\label{potential2}
\big[\EE+k_0^2\Delta\tilde\epsilon(\rrh)
+\nabla_\rrh^2\big]E_\alpha(k_z,\rrh,\omega)&=&0\qquad\mbox{PT}\nonumber\\
\big[\EE+\nabla_\rrh\frac{1}{\tilde\epsilon(\rr)}\nabla_\rrh\big]B_\alpha(k_z,\rrh,\omega)&=&0\qquad\mbox{MT}
\ea
with
$\rrh=x{\bf e}_x+y{\bf e}_y$
and $\Delta\tilde\epsilon(\rrh)=\tilde\epsilon(\rrh)-1$.
Here we have introduced the spectral parameter (eigenvalue)
\mbox{$\EE=k_\perp^2=k_0^2-k_z^2=k_0^2\sin(\theta)^2$,}
where $\theta$ is
the azimuthal angle.

\begin{figure}
\includegraphics[width=8cm,clip=true]{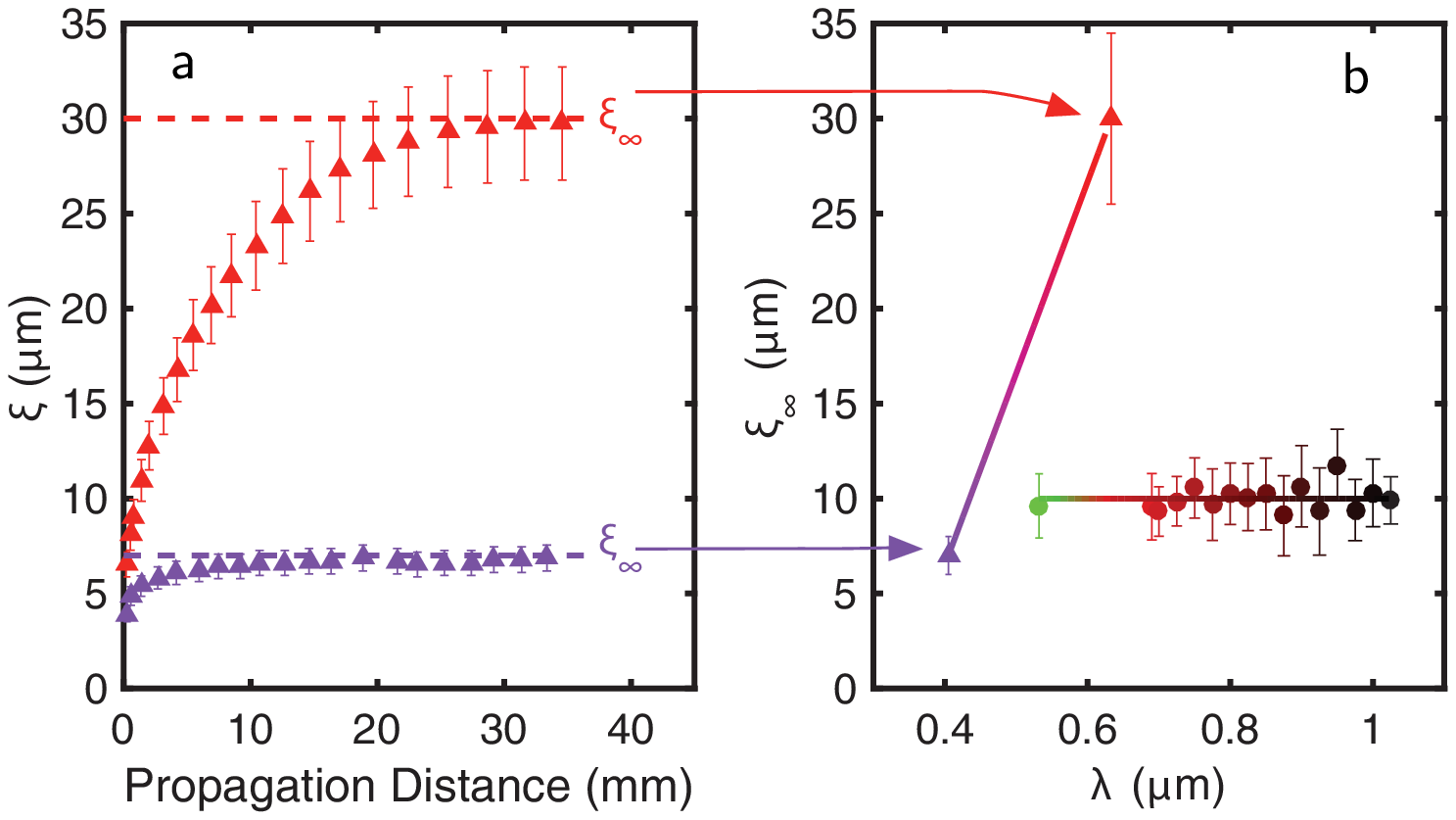}
\caption{
a:
Localisation radius $\xi(z)$
as a function of the distance $z$ along the
fibre
for the two light wavelengths
$\lambda =0.4\,\mu$m and $\lambda=0.63\,\mu$m from the simulation of Karbasi%
\cite{karbasi12a} based on the potential-type version
of Eqs. (\ref{potential3}).\newline
b: Measured averaged
localization length of fibres with transverse
disorder as a function of the incident-laser wavelength
(full circles), compared
with the two values $\xi_\infty\equiv\xi(z=\infty)$ of panel a
(full triangles).
}
\label{marco}
\end{figure}

We note that in the PT equation the wavenumber $k_0=2\pi/\lambda$
appears as an external parameter in front of the fluctuating permittivity,
whereas in the MT version $k_0$ enters only via the spectral
parameter $\EE$.

In the paraxial limit
$\theta\rightarrow 0$ $\EE$ can approximated
by $-2k_0\Delta k_z$, where $\Delta k_z=k_z-k_0$ is the
Fourier wavenumber corresponding to the $z$ dependence
of the {\it envelopes}
$E_\alpha^{(0)},
B_\alpha^{(0)}$
of the electromagnetic fields.
Transformed
back to the $z$ dependence of the envelope one obtains
\ba\label{potential3}
\big[2ik_0\frac{\partial}{\partial z}+k_0^2\Delta\tilde\epsilon(\rrh)
+\nabla_\rrh^2\big]E_\alpha^{(0)}(z,\rrh,\omega)&=&0\quad\mbox{PT}\nonumber\\
\big[2ik_0\frac{\partial}{\partial z}+\nabla_\rrh\frac{1}{\tilde\epsilon(\rr)}\nabla_\rrh\big]B^{(0)}_\alpha(z,\rrh,\omega)&=&0\quad\mbox{MT}
\ea
Eqs. (\ref{potential3}) are
mathematically equivalent to a time-dependent
Schr\"odinger equation (with ``time'' $z/2k_0$).

The PT version of Eqs. (\ref{potential3})
has been used in
Refs. \cite{deraedt89,schwartz07,karbasi12,karbasi12a} for a numerical
calculation of the localization properties of transverse-disordered
optical fibres.
In the panel a of Fig. 1 we have reproduced the $z$
dependence of the radius of the localization lengths $\xi(z)$, obtained by
such a simulation\cite{karbasi12a}
for two different light wavelengths $\lambda$,
wich saturate for large
$z$ at the localization length $\xi(z\!=\!\infty)\equiv\xi_\infty$. The strong dependence
on the wavelength is striking.

In order to check, whether this behaviour predicted by the PT
theory is realistic,
we have taken samples
composed of microfibres with different permittivity
(polystyrene, PS $\epsilon_{\rm PS}/\epsilon_0=1.59$,
polymethylmethacrylate, PMMA, $\epsilon_{\rm PMMA}/\epsilon_0=1.49$),
fabricated as described in \cite{karbasi12}
in which transverse localization is observed%
\cite{karbasi12,karbasi12a,leonetti14,mafi-oe}.
We measured the localization length in such devices
by injecting a focused (order of a
micrometer) monocromatic light at the fibre input tip while
monitoring the total fibre output.
The average extent of the output intensity pattern is determined by the
localization length in the fibre.
Thus we estimated it by determining
the first spatial moment of the intensity distribution.
Averaging is performed by scanning the input facet with a 3-axis piezo motor sustaining the fibre. The experimental setup together with a sketch of the
fibre geomety is shown in Fig. 2. A more detailed description of our
experiment can be found in the supplemental material.

In panel b
of Fig. 1 we show our data for the localization length $\xi_\infty$
(full circles),
averaged over all modes and three samples as a function of the
incident laser wavelength. It can be seen that there is no change
with the wavelength.

Our interpretation is that this discrepancy
is due to the inadequateness of the potential-type stochastic wave
equation.

\begin{figure}
\includegraphics[width=8cm,clip=true]{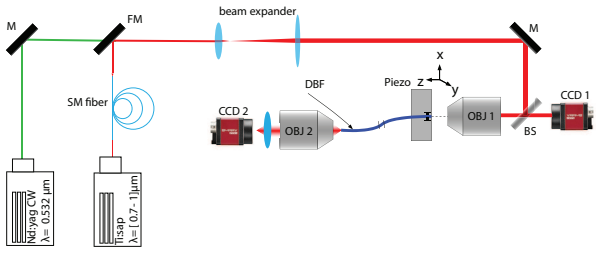}
\caption{Sketch of the experimental setup: The light
from a ND:YAG or a Ti:Sapphire laser is coupled to the
fibre by using objective OBJ 1.
The back-reflected light is then visualized by the camera CCD1
through the beamsplitter BS in order to perform the fine tuning of the alignment.
The piezo devices control the laser injection location.
The transmitted light is collected by the objective OBJ2 ad imaged on camera CCD2 with a magnification of 50. \newline
In (a)  a sketch of the fibre is reported,
while in (b) a magnified image of the fibre tip surface is shown,
where polymethylmethacrylate appears dark and polystyrene white. }
\label{setup}
\end{figure}

But how come that the results of the two descriptions, which are both supposed
to arise from Maxwell's equation with spatially varying permittivity,
differ from each other? For deriving the wave equations in the presence
of an inhomogeneous permittivity $\epsilon(\rr)$ one can either
solve for the electrical field $\Ee(\rr,t)$ (PT) or divide through
$\epsilon(\rr)$ and then solve for the magnetic field $\BB(\rr,t)$
(MT).
In the first case
one has to decompose the double curl of $\Ee$, whereas
in the second case the double curl of $\BB$:

\ba\label{wave2}
\frac{\epsilon(\rr)}{c_0^2}
\frac{\partial^2}{\partial t^2}\Ee(\rr,t)
&=&-\nabla\times\nabla\times\Ee(\rr,t)\qquad\mbox{PT}\nonumber\\
&=&\nabla^2\Ee-\nabla\big(\nabla\cdot\Ee(\rr,t)\big)\nonumber\\
\frac{\partial^2}{\partial t^2}\BB(\rr,t)
&=&-\nabla\times
\frac{c_0^2}{\epsilon(\rr)}
\nabla\times\BB(\rr,t)\qquad\mbox{MT}\nonumber\\
&=&\nabla\cdot \frac{c_0^2}{\epsilon(\rr)}\cdot \nabla \BB(\rr,t)
\ea

In the absence of free charges but in the presence
of a spatially fluctuating dielectric constant
we get for the divergence of the electic field
\be
\nabla \cdot\Ee=
-\frac{1}{\epsilon_0}
\nabla \cdot\PP=
-\frac{1}{\epsilon_0}
\nabla\cdot[\epsilon(\rr)-\epsilon_0]\Ee
\ee
from which follows\cite{saleh}
\be
\nabla\cdot\Ee=-\frac{1}{\epsilon(\rr)}\Ee\cdot\nabla\epsilon(\rr)\neq 0
\ee
The divergence of $\BB$, on the other hand, is zero.

Obviously, in the first paper using the PT approach\cite{john84} and
the whole following literature\cite{john85,john87,sheng06,deraedt89,%
schwartz07,karbasi12,karbasi12a,mafi-prb}
the
divergence of $\PP$ (which describes the frozen-in
displacement charges)
had been tacitly assumed to be zero%
\footnote{with the exception of Ref. \cite{mafi-josa}}.
We believe that this is the origin of the discrepancy of the
two theories.

We further check the consistency or otherwise of the two
approaches
by applying the theory of the nonlinear
sigma model of localization to the stochastic Helmholtz
equations (\ref{potential2}). Wegner\cite{wegner79} realized
that the nonlinear sigma model of planar ferromagnetism
obeys the same scaling of the
coupling constant with the length scale $L$
as the conductance $g$ of electrons
in the scaling theory of localization\cite{abrahams79}, namely
\be\label{scaling}
\frac{d g}{d\ln L}=g\beta(g)=(d-2)g-c
\ee
where $d$ is the dimensionality and
$c$ is a constant of order unity.
Later a rigorous mapping
of the field-theoretical representation of the
configurationally averaged Green's functions
to a generalized
nonlinear sigma model was established\cite{schafer80,houghton80,mckane81}.
This was then
adapted to classical sound waves\cite{john83,schirm06},
and -- using the PT approach -- to electromagnetic waves%
\cite{john84,john85,john87}.
For $d=2$
the solution of Eq. (\ref{scaling}) is
\be
g(L)=g(L_0)-c\ln(L/L_0)
\ee
where $L_0$ is the reference length scale, i.e. $g$ scales always towards zero.
The localization length
$L\equiv \xi_\infty$ is the length at which $g=1$\cite{mckane81,john83},
and $g(L_0)=g_0$ is the reference conductance.

The nonlinear-sigma-model theory provides us
via a saddle-point approximation with
a nonperturbative way to calculate the reference conductance, which
is related to the scattering mean-free path $\ell_0(\EE)$.
Within
this saddle-point approximation
(self-consistent Born approximation, SCBA%
\cite{john83,schirm06})
$g_0(\EE)$ and $\ell_0(\EE)$ are given in terms of a complex
self-energy function $\Sigma(s)=\Sigma'(\EE)+i\Sigma''(\EE)$
with complex spectral parameter $s=\EE+i\eta,\eta\rightarrow 0$
\ba\label{ioffe1}
\ln\xi_\infty(\EE)\propto g_0(\EE)&=&
\frac
{\EE+k_0^2\Sigma'(\EE)}
{k_0^2\Sigma''(\EE)}\qquad\quad\,\,\,\mbox{PT}\nonumber\\
\ln\xi_\infty(\EE)\propto g_0(\EE)&=&
\frac{
1-\Sigma'(\EE)
}{
\Sigma''(\EE)
}\qquad\quad\mbox{MT}
\ea
The function $\Sigma(s)$
obeys the self-consistent equation
\ba\label{scba1}
\Sigma(s)&=&\gamma\frac{k_0^2}{q_c^2}
\int_0^{q_c}\!\!q\,dq\,\Gg(q,s)\qquad\qquad\mbox{PT}\nonumber\\
\Sigma(s)&=&
\gamma
\frac{1}{q_c^2}
\int_0^{q_c}\!\!q\,dq\,q^2
\Gg(q,s)\qquad\qquad\mbox{MT}
\ea
with the disorder parameter $\gamma=\langle(\Delta\tilde\epsilon)^2\rangle$
and the averaged one-particle Green's functions
\ba\label{green1}
\Gg(q,s)
&=&\frac{1}{-s-k_0^2\Sigma(s)+q^2}\equiv\frac{1}{-k_\Sigma(s)^2-q^2}%
\,\,\mbox{PT~~~}\nonumber\\
\Gg(q,s)
&=&\frac{1}{-s+q^2[1-\Sigma(s)]}\equiv
\frac{[1-\Sigma(s)]^{-1}}{-k_\Sigma(s)^2-q^2}\,\,\mbox{MT~~~}
\ea
where we have introduced a complex wavenumber
$k_\Sigma(s)=k_\Sigma'(\EE)+ik_\Sigma''(\EE)$ in
an obvious way. The imaginary part of this
quantity is related to the scattering mean-free path
by $\ell_0(\EE)=1/2k_\Sigma(\EE)''$
and we obtain for both descriptions (see \cite{john83,schirm06} and
the supplementary material)
\be
g_0(\EE)=
k'_\Sigma (\EE)\ell_0(\EE)
\ee
The upper $q$ integration limit is given by the correlation parameter
$q_c=2\pi/L_c$%
\cite{john83a,tomaras08}, where $L_c$ is the
disorder correlation
length ($\sim $ diameter of the grains with different permittivities).

\begin{figure}
\includegraphics[width=8cm,clip=true]{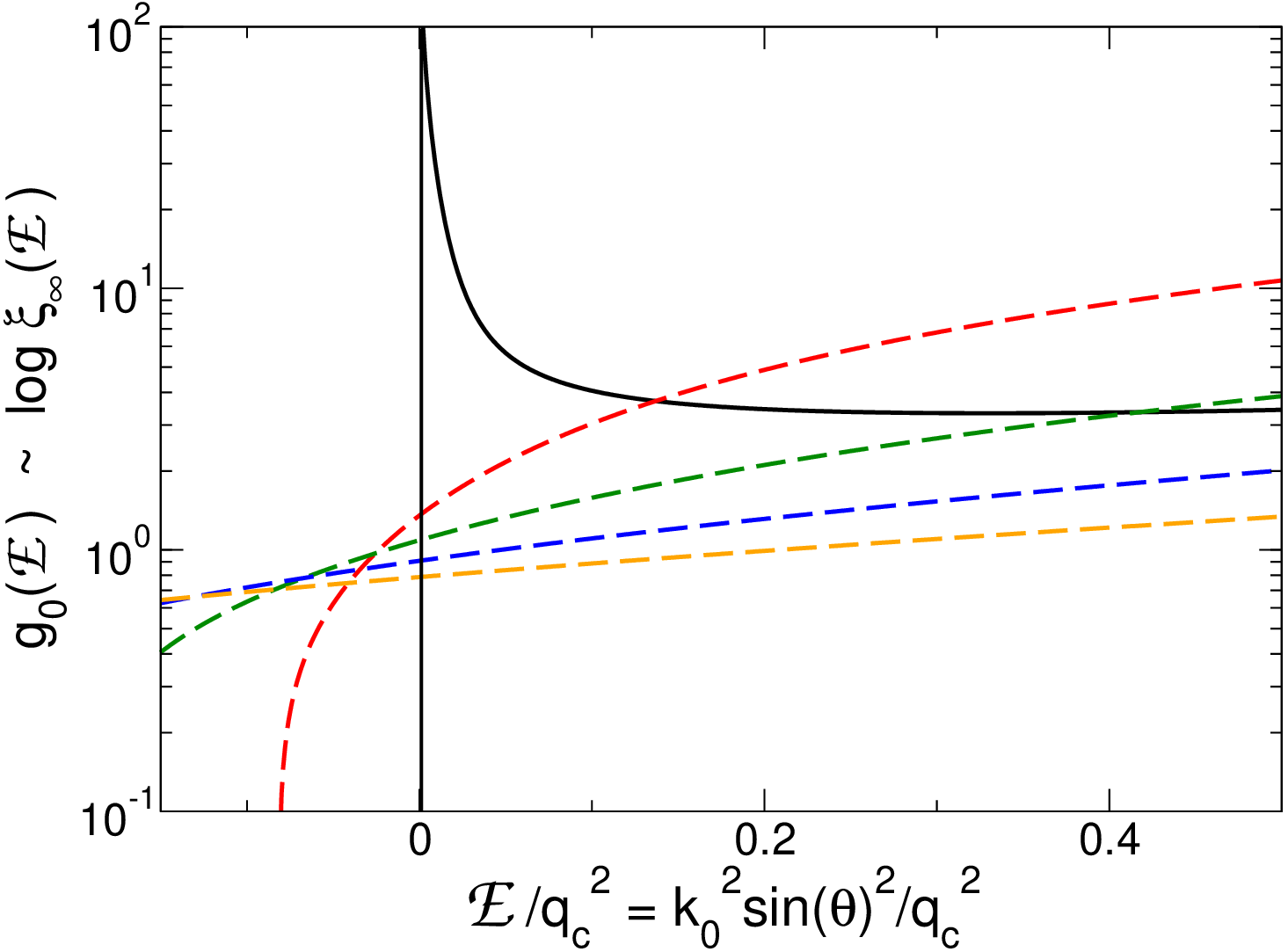}
\caption{``Conductance'' $g_0(\EE)\propto \ln{\xi_\infty(\EE)}$,
where $\xi_\infty(\EE)$ is the localization lenght,
for the modulus-type description (continuous line) and for
the potential-type description (dashed lines)
with four different wavelengths
$\lambda=1\mu{\rm m},
\lambda=0.75\mu{\rm m},
\lambda=0.6\mu{\rm m},
\lambda=0.5\mu{\rm m}$.
For both calculations we used the disorder parameter $\gamma$ = 0.2.
The correlation parameter $k_c$ has been determined from the
spatial distribution of the dielectric constants, see
Fig. 2,
by an
image-processing correlation
analysis to be $k_c=8\,[\mu{\rm m}]^{-1}$.
}
\label{conductance}
\end{figure}

In Fig. 3 we show the
reference conductance $g_0(\EE)$,
which is proportional to the logarithm of the localization
length $\xi_\infty(\EE)$,
for the two alternative theories
against the spectral parameter $\EE$.
As to be expected, the $g(\EE)$ curves
depend strongly on $k_0$ for the PT model, in contrast to
the MT case, in which the $g(\EE)$ curves do not
depend on the wavelength. In this case
the only dependence on $k_0$ is via
the spectral parameter $\EE$. From this it follows that
in the MT case the distribution of localization lengths
(and hence their average) do not depend on the wavelength,
in agreement with our data displayed in Fig. 1b and in
disagreement with the numerical (Fig. 1a) and field-theoretical
(Fig. 3) predictions of the PT theory.

As reference length scale $L_0$ we use the disorder correlation
length $L_c$ and not the scattering mean-free path\cite{lr85,vollhardt80}.
With $L_c\approx$ 1$\mu$m
this is consistent with our measured values of $\xi_\infty\approx$ 10 $\mu$
corresponding to $g_0\approx 5$, which is obtained near
$\EE/q_c^2\approx 0.07$. This value of $\EE$ is well inside the
maximum value of $[\EE/q_c^2]_{\rm max}=0.5L_c^2/\lambda^2$ of our
device, determined
by $\theta_{\rm max}=50^{\circ}$.

Within the modulus-type description
the reference
conductance diverges
as $\EE^{-1}$ for $\EE\rightarrow 0$. For the mean-free path
one obtains
$\ell_0(\EE)\propto \EE^{-3/2}$, which is equivalent to a
two-dimensional Rayleigh
law. This absence of scattering for rays
entering the sample exactly in the $z$ direction
indicates that the sample is lengthwise transparent, as
it should. The Rayleigh law can also be written
as $\ell_0\propto (\lambda_\perp/L_c)^3$, where
$\lambda_\perp=\lambda/\sin \theta$ is the transverse
wavelength. So if $\lambda_\perp$ becomes much larger than
the grain size $L_c$, there is no scattering and hence
no localization.

On the contrary,
within the potential-type description the sample is
predicted to be opaque in the $z$ direction, and
the Rayleigh law is absent.
It is even seen from Fig. 3, that the spectral scattering intensity
of the potential-type model extends into the negative $\EE$ range,
rendering the spectrum unstable. Stability, i.e. the restriction
of the spectrum to positive values, is however required for
disordered bosonic systems%
\cite{lueck06}.

We see that the previous theoretical approach, the potential-type
one, leads to general inconsistencies and, in particular, to
an inconsistency with our measured localization data. This is not the case
with the modulus-type description, which leads to a positive spectrum,
predicts transparency in the $z$ direction and is consistent with
our mesured data on transversely localized optical devices.
Thus we are convinced that we have now established a sound theoretical
fundament for further work on Anderson localization of light.

\end{document}